# Highly-Accurate Electricity Load Estimation via Knowledge Aggregation

Yuting Ding, Di Wu, *Member*, *IEEE*, Yi He, *Member*, *IEEE*, and Xin Luo, *Senior Member*, *IEEE*

Song Deng, *Member*, *IEEE*

*Abstract*—Midterm and long-term electric energy demand prediction is essential for the planning and operations of the smart grid system. Mainly in countries where the power system operates in a deregulated environment. Traditional forecasting models fail to incorporate external knowledge while modern data-driven ignore the interpretation of the model, and the load seires can be influenced by many complex factors making it difficult to cope with the highly unstable and nonlinear power load series. To address the forecasting problem, we propose a more accurate district level load prediction model Based on domain knowledge and the idea of decomposition and ensemble. Its main idea is three-fold: a) According to the non-stationary characteristics of load time series with obvious cyclicality and periodicity, decompose into series with actual economic meaning and then carry out load analysis and forecast. 2) Kernel Principal Component Analysis(KPCA) is applied to extract the principal components of the weather and calendar rule feature sets to realize data dimensionality reduction. 3) Give full play to the advantages of various models based on the domain knowledge and propose a hybrid model(XASXG) based on Autoregressive Integrated Moving Average model(ARIMA), support vector regression(SVR) and Extreme gradient boosting model(XGBoost). With such designs, it accurately forecasts the electricity demand in spite of their highly unstable characteristic. We compared our method with nine benchmark methods, including classical statistical models as well as state-of-the-art models based on machine learning, on the real time series of monthly electricity demand in four Chinese cities. The empirical study shows that the proposed hybrid model is superior to all competitors in terms of accuracy and prediction bias.

## I. Introduction

Maintaining a continuous balance between power consumption and production is prerequisite for the stable and efficient operation of the power system. Meanwhile, it poses a serious chanllenge which has been exacerbated in recent years by the growing share of renewable energy[1]. The key requirement of balancing power system is that the demand and generation capacity at any point is reliably predicted in time. For example, a study by the California Energy Commission shows that improved solar energy and load forecasting can generate potential savings of USD 2 million yearly[2]. Therefore, accurate prediction of power supply and demand is of great significance not only to ensure the safe and efficient operation of the system, but also to increase market returns and reduce financial risks.

Currently, numerous papers have focused on monthly electricity consumption forecasting methods[3]–[5]. The monthly load demand forecasting methods can be generally classified as statistical/econometric models or machine learning(ML)/computational intelligence models[6]. Although statistical methods have improved the accuracy of load forecasting to some extent, some recent studies cannot satisfy the assumption of linearity of electric loads, thus limiting the accuracy of forecasting[7]–[9].

In order to reduce the influence of strong non-stationary components generated by wavelet packet decomposition on prediction, the ensemble empirical mode decomposition is used to decompose this component again in literature[10-13]. However, the ensemble empirical mode decomposition introduces noise, and the components generated by the second decomposition are numerous, which will cost a lot of time to predict by neural network. In this regard, in literature [14-16], the non-stationary components generated by ensemble empirical mode decomposition are predicted by gated cyclic units, and the stationary components are predicted by multiple linear regression (MLR)[24], which saves a lot of time cost. However, it is also difficult to accurately predict the high-frequency components after the introduction of noise and decomposition. Guo et al. considered the impact of economic disturbance on electricity consumption and used the X12-ARIMA model combined with economic factors and load series to forecast the electricity demand[17-20]. In the literature[13] decomposed the time series into trend and seasonal components using a classical decomposition model. The time series was decomposed into trend and seasonal components and seasonal autoregressive integrated moving average (SARIMA) and weighting methods were used to predict each component. Nevertheless, It is only a simple weighted prediction component and does not combine the relevant influencing factors, the component prediction accuracy is not high.

To the authors' best knowledge, Most of the existing decomposition and aggregation modelscannot take into account the statistical properties of load sequences such as periodicity, instability, and external factors affecting load variation at the same time like RD-ETS+LSTM and NBeats[21], [22]. Related studies have also proposed combining complex factors to predict load sequences, rather than relying solely on algorithmic models[23-28]. What would be the accuracy of the prediction model if the complex factors are combined to predict for load characteristics? Based on this key question Motivated by these critical issues, this paper proposes a accurate electricity load estimation via knowledge aggregation. Regarding electricity consumption data and external factors selected from feature engineer, on the basis of seasonal adjustment through Census X12 decomposition method, A hybrid approach including ARIMA, SVR and XGBoost is used to forecast the electricity consumption. Therefore, it is significantly different from existing models. It achieves both robustness and stability to forecast the district electricity consumption. Main contributions of this study include:

a) We propose a prediction algorithm based on ARIMA and SVR, which effectively takes into account the statistical characteristics of the load and the influence of external factors on it, so as to improve the prediction accuracy of the load components.

b) Considering the characteristics of the power load sequence itself and removing the influence of its internal factors, we propose to decompose the load to forecast separately to improve the forecast accuracy.

c) Algorithm design and analysis for an XASXG model.

A careful empirical study of real data sets from four cities in China is conducted to evaluate the performance of XASXG. The results show that the XASXG model achieves significant accuracy gains in forecasting monthly electricity compared to the state-of-the-art electricity load forecasting model first. Its accuracy is also higher compared to other forecasting models.

## III. FORECASTING MODEL

### A. Motivation of XASXG

By studying the laws of the electricity sequence it can be found that electricity changes with the development of time, such as seasonal changes leading to an increase or decrease in electricity consumption power consumption, the month contains not only social behavior information such as holidays also contains natural information represented by the seasons. In addition to being related to the month, the turning point of the electricity consumption curve is also related to the sudden change of the season. In addition, we all know that economic development requires the development of electricity, and the development of regional economy is closely related to the demand for electricity. Fig. 1 visualizes the original electric time series and its components. Therefore, due to the different influencing factors, the electricity time series can be decomposed into series containing different characteristics, such as trend components that are mainly influenced by economic factors and reflect the direction of development over a longer period of time; seasonal components that are influenced by seasonal changes and have cyclical fluctuations; and random components that are influenced by various chance factors. Based on this, it is necessary to reduce the influence of complex factors on the forecasting model while combining external knowledge to forecast the final aggregation of each component separately.

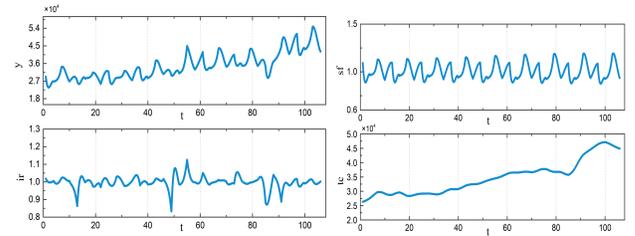

Fig. 1. The original electric time series and its components.

### B. Framework and Features

The proposed forecasting model is shown in Fig. 2. It is composed of the following.

1) Load Composition—X12 The decomposition technique can decompose periodic time series such as economic volume and monthly electricity into trend components $Tc$, seasonal components $St$ and irregular components $Ir$, The multiplication model assumes that the influence of each component on the development of phenomena is interrelated based on the absolute amount of the trend component, and the other components are expressed in proportion[25-29]. Besides, a multiplicative model is applied for seasonal decomposition in this article.

2) Feature Engineer—The components used in predicting monthly load can be divided into historical data, climate variables, calendar variables, and holiday data. In this paper, Pearson Correlation Coefficient and Feature importance test are used to pick out features that are more relevant to the monthly electric load.

3) Hybrid Model Construction— Trend components, seasonal components, and irregular components are the directions of development formed by the influence of different components. In this paper, the corresponding prediction models are constructed according to the statistical characteristics of different components[30-32]. The ARIMA-SVR model is used to predict the trend element, and the XGBoost model to predict the seasonal component and the irregular component. Finally, the three components prediction results are multiplied and aggregated into monthly loads.

4) Evaluation of prediction model— In this paper, the prediction models used are evaluated from four perspectives: accuracy, generation, stability, and Applicability.

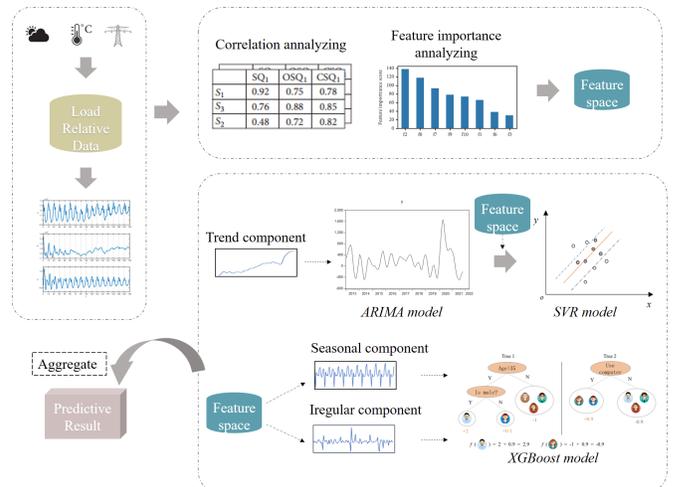

Fig. 2. A general framework of XASXG

## C. Feature engineer

the correlation analysis is a statistical test on whether the variables are related and the degree of correlation. We use the Pearson correlation coefficient to analyze the correlation between the features. The value of the Pearson correlation coefficient is between -1 and 1. This linear correlation is expressed intuitively as whether $Y$ increases or decreases at the same time as $X$ increases; when the two are distributed on a straight line, the Pearson correlation coefficient is equal to 1 or -1; when there is no linear relationship between the two variables, the Pearson correlation coefficient is 0.

On the other hand, features are important to the model, but some features can lead to redundancy, and it will interference with model prediction accuracy. Therefore, we should filter features. Based on the scores of feature importance, the higher the score is, the more important the features are, and the features with lower scores can be discarded. Finally, the features with strong correlation and high importance scores are retained as the feature space of the prediction model.

## D. The predictive model of three components

### 1) Trend component forecast

For prediction model construction, we then aim to determine an appropriate training model for each component. First, the patterns and characteristics of the different components are analyzed for prejudgement of model selection. The trend element is a long-term trend rising significantly with time, thus the ARIMA-SVR model is considered to use in this paper. we consider modeling it using ARIMA-SVR.

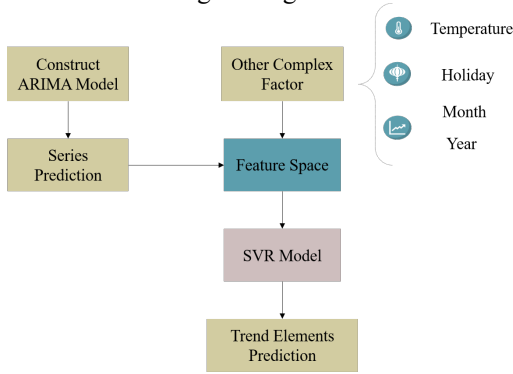

Fig. 3. The flow chart of the ARIMA-SVR

The ARIMA model is based on the statistical learning theory, which comprehensively considers the fitting error and function characteristics of the regression model, which ensures that it has good generalization ability; and the SVR model is based on kernel mapping, which ensures its nonlinear processing ability. Besides, SVR has achieved good prediction results in the field of nonlinear time series, and solved the problems of small sample, nonlinearity, over-fitting, dimension disaster and local small, etc., with excellent generalization ability. At the same time, the ARIMA-SVR model combines the statistical characteristics of the electricity load series itself and considers the influence of other complex factors on the load. The flow chart of the ARIMA-SVR model is shown in Fig. 3.

### 2) Seasonal component and irregular component forecast

Generally, the seasonal cycle component of the monthly load series shows a series of variables with relatively small volatility and has a certain variation pattern. It is seen that the seasonal component value of the current month does not change much from the value of the same month last year, so this paper the value of the same month last year with temperature, holidays, months, as the feature space of prediction model in the seasonal component prediction. The random chance component of the monthly load has no obvious change pattern, but it is related to some chance factors including earthquakes, severe weather, Chinese New Year, etc. The occurrence of these chance factors will certainly lead to a sudden increase or sudden decrease of monthly electric load. Therefore, combined with feature engineering, temperature, holidays, and months are considered as the feature space of the prediction model. After selecting the features, we can build the model. XGboost model is considered to train the model and use the 10-fold cross-validation method to verify the model during the training process. An overall process of building Finally, the built XGBoost is used to predict monthly electric load.

## IV. EXPERIMENTS AND RESULTS

### A. General Settings

### 1) Datasets

We select the monthly electricity load from four citys in china. The time series have different lengths: nine years( HM, QD). The time series are present in Fig.1. As we can see from this figure, the monthly electricity load series exhibit nonlinear trends, a strong seasonal cycle, and variability.

### 2) Evaluation Metrics

We are mainly concerned with the prediction accuracy and efficiency of the model under test. For prediction accuracy, we are interested in how close the estimates are to the actual elements, as it directly reflects whether the model captures the essential features of the given data [33,34]. mean absolute percentage error (MAPE) is often chosen as metrics to assess accuracy. MAPE can reflect the reliability of the predicted result. MAPE can not only measure the deviation between the predicted result and the actual value but can also consider the ratio between the error and the actual value. the lower the MAPE, the better the model. Mean absolute error (MAE) is another loss function used in regression models. the MAE is the sum of the absolute values of the differences between the target and predicted values.

$$MAPE = \frac{1}{N}\sum_{i=1}^{N}\left|\frac{y_i - \hat{y}_i}{y_i}\right|\times 100\% \qquad (22)$$

$$MAE = \frac{1}{m}\sum_{i=1}^{m}|(y_i - \hat{y}_i)| \qquad (23)$$

### 3) Baselines

We compare XASXG with nine related models with different characteristics, including five ML-based models (GRNN[16], XGBoost[19] , MLP[24], LSTM[38],), three statistic models (ARIMA[26], X12-ARIMA[24], ETS[37]), and two state-of-the-art models(RD-ETS+LSTM[1],NBEATS[3]).

*4) Experimental Designs*

The last 12 months of data are used as the test set and the previous data as the training set. The forecast model and comparative model were used to forecast the monthly electricity load from 12/2020 to 11/2021. The data from previous years were used for hyperparameter selection and learning. Specifically, for the selection of hyperparameters, the models were trained on the training set. Then, the selected optimal hyperparameters were used to construct the model for the test set. The selected hyperparameters were as follows:

1) Learning rate in XGBoost: 0.02
2) Number of estimators in XGBoost: 1000
3) Number of autoregressive terms: $p = 2$
4) Moving average number of terms: $q = 1$
5) Difference order: $d = 2$

The results reported in this article take averages over 30 independent trials (learning sessions). The model was implemented in python relying on the Sklearn library. We run the experiments on a personal computer (Intel i7 CPU 3.4 GHz, RAM 16 GB).

*B. Comparison of rating prediction accuracy*

**Checking the prediction accuracy of XASXG** Fig. 4 shows the predictive result of different models for each citys. As can be seen from this figure, XASXG is the most accurate model in most cases. Note that the error in the 1st-4th month is small, while that in the 6th-8th month and 12th month is larger. The result of XASXG is better than that of most comparison models, and the fitting effect of XASXG in HM data set is the best. Due to the dataset chosen by the experiment, the classical method fits the load series better than the ML method. But ETS+RD-LSTM and XASXG which combine the ML and classical methods work better than classical methods alone. In addition, it is most obvious that the predictive power of the MLP plays the worst role in these four datasets, with the worst sensitivity to different months. From Fig. 5, we can find that the prediction accuracy of the XGBoost model fluctuates greatly, and its prediction ability is sometimes very accurate and sometimes very wrong in different months. Rankings of the models are shown in Fig. 4. These are based on average ranks of the models in the rankings for individual cities. The top of Fig. 4 shows the ranking based on MAPE and the bottom of Fig. 4 shows the ranking based on MAE. Note the high position of XASXG. It is in the first position in MAPE ranking and the fourth position in MAE ranking. Note that in the latter case, the difference between the first four positions is very small.

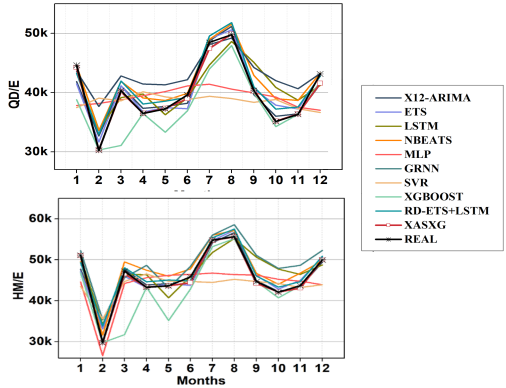

Fig. 4. Real and forecast monthly electricity demand for selected citys

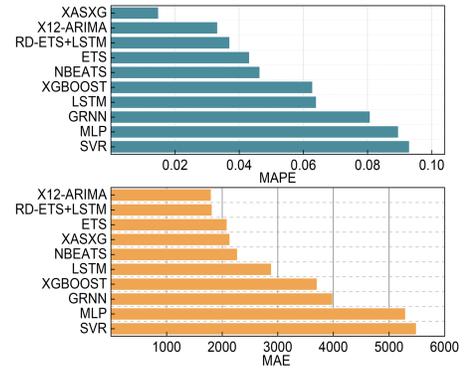

Fig. 5. Rankings of the models

TABLE I
THE COMPARISON RESULTS ON PREDICTION ACCURACY, INCLUDING WIN/TIE/LOSS COUNTS STATISTIC AND FRIEDMAN TEST, WHERE • INDICATE THAT THE PREDICTION ACCURACY OF XASXG IS HIGHER THAN OR SAME TO THAT OF COMPARISON MODELS RESPECTIVELY.

| Month | X12-ARIMA | ETS | LSTM | NBEATS | MLP | GRNN | SVR | XGBOOST | RD-ETS+LSTM | XASXG |
|---|---|---|---|---|---|---|---|---|---|---|
| 1 | 6.60% | 7.83% | 1.00% | 3.08% | 12.69% | 2.23% | 14.84% | 8.40% | 3.59% | 0.12% |
| 2 | 13.48% | 11.21% | 14.93% | 6.15% | 10.67% | 18.81% | 19.53% | 0.11% | 13.69% | 0.01% |
| 3 | 0.95% | 2.61% | 1.69% | 4.76% | 6.41% | 3.38% | 5.39% | 32.80% | 1.87% | 1.50% |
| 4 | 1.33% | 0.21% | 7.07% | 9.64% | 5.48% | 12.40% | 7.57% | 0.01%• | 3.11% | 0.18% |
| 5 | 1.24% | 1.17% | 6.67% | 5.37% | 5.93% | 1.21% | 2.53% | 19.42% | 3.28% | 0.03% |
| 6 | 4.46% | 4.44% | 0.29% | 4.22% | 1.39%• | 5.60% | 2.40%• | 6.45% | 2.00% | 3.10% |
| 7 | 1.15% | 0.17% | 5.54% | 2.14% | 14.69% | 2.26% | 18.81% | 2.89% | 1.07% | 0.16% |
| 8 | 1.75% | 2.82% | 0.76% | 2.69% | 16.66% | 5.28% | 18.70% | 0.91% | 3.40% | 0.06% |
| 9 | 0.44% | 3.30% | 13.34% | 4.47% | 3.48% | 14.28% | 0.05%• | 0.01%• | 2.95% | 1.09% |
| 10 | 0.38% | 1.75% | 13.36% | 4.66% | 7.50% | 13.75% | 5.01% | 3.30% | 2.97% | 0.10% |
| 11 | 0.67%• | 2.94% | 6.27% | 7.06% | 2.51% | 11.41% | 0.96%• | 0.00%• | 2.20% | 1.38% |
| 12 | 1.00% | 1.50% | 2.43% | 0.40% | 12.07% | 4.65% | 12.09% | 0.21% | 1.71% | 0.03% |
| **Win/Loss•** | 11/1 | 12/0 | 12/0 | 12/0 | 11/1 | 12/0 | 9/3 | 9/3 | 12/0 | — |
| **F-rank** | 4 | 4.833 | 6.083 | 6 | 7.166 | 7.666 | 7.25 | 5 | 5 | **2** |
| **P-value** | 0.0080 | 0.0002 | 0.0024 | 0.0002 | 0.0007 | 0.0002 | 0.003 | 0.0212 | 0.0017 | — |

* The smaller F-rank value denotes a higher rating prediction accuracy on HM.

**Statistical analysis.** According to the actual tests of the above two groups of real data, the comparison results with other models in different months show that our method is generally better than or close to the comparison model. In addition, in order to verify whether the XASXG model has better data recovery performance than other models, we make a statistical analysis of each table. First of all, the win/lose/win situation of this model and other comparative models is summarized from the third row to the last row of each table, which shows that the XASXG model is superior to other models in most data sets. Secondly, we conduct Friedmantest tests to verify the recovery performance of the model on multiple datasets. The recovery performance of the model on multiple datasets. The MAE/MAPE test results for each table are recorded in the last row of each table. The MAE/MAPE test results for each table are recorded in the last row of each table. The test results show that in the penultimate row of each table, it is better than other models. The Wilcoxon signature rating test is a way to check whether our model has better data recovery performance than other models. In the penultimate row of each table, the Wilcoxon signature rating test is to check whether our model has better performance[35,36]. The P value is less than 0.05, which indicates that our model has better performance. The model has higher performance. Summary: it can be seen from the table I-II that XASXG has made progress in forecasting the components of different months, so the total amount of electricity aggregated is basically close to the real value.

TABLE II
THE COMPARISON RESULTS ON PREDICTION ACCURACY, INCLUDING WIN/TIE/LOSS COUNTS STATISTIC AND FRIEDMAN TEST, WHERE • INDICATE THAT THE PREDICTION ACCURACY OF XASXG IS HIGHER THAN OR SAME TO THAT OF COMPARISON MODELS RESPECTIVELY.

| Month | X12-ARIMA | ETS | LSTM | NBEATS | MLP | GRNN | SVR | XGBOOST | RD-ETS+LSTM | XASXG |
|---|---|---|---|---|---|---|---|---|---|---|
| 1 | 5.77% | 7.15% | 1.33% | 5.45% | 5.19% | 3.20% | 8.42% | 10.32% | 3.11% | 0.05% |
| 2 | 18.13% | 15.03% | 8.00% | 13.23% | 55.47% | 14.75% | 65.68% | 0.31% | 16.92% | 0.02% |
| 3 | 1.49% | 3.83% | 5.12% | 5.35% | 4.81% | 17.15% | 0.25% | 33.50% | 1.97% | 0.05% |
| 4 | 1.60% | 2.77% | 1.61% | 6.37% | 5.53% | 7.92% | 6.71% | 0.00% | 2.43% | 0.43% |
| 5 | 1.40% | 1.22% | 4.19% | 4.32% | 2.01% | 4.04% | 3.75% | 17.17% | 2.50% | 0.04% |
| 6 | 3.19% | 3.91% | 1.26% | 0.34% | 1.75% | 1.64% | 0.12%• | 4.91% | 1.21% | 1.64% |
| 7 | 1.04%• | 0.55%• | 0.35% | 0.96% | 11.66% | 5.80% | 12.68% | 9.95% | 0.78% | 1.52% |
| 8 | 0.77%• | 2.16% | 10.08% | 7.28% | 16.05% | 4.82% | 14.22% | 4.75% | 1.33% | 1.11% |
| 9 | 5.40% | 8.48% | 20.10% | 5.72% | 6.07% | 8.07% | 6.14% | 0.01%• | 6.89% | 0.16% |
| 10 | 1.80% | 3.77% | 11.37% | 6.32% | 6.08% | 7.04% | 7.00% | 1.50% | 3.88% | 0.04% |
| 11 | 0.58% | 2.37% | 5.30% | 3.32% | 0.20% | 5.21% | 0.65% | 0.00%• | 2.77% | 0.01% |
| 12 | 16.99%• | 12.59%• | 9.99%• | 8.19%• | 24.56% | 14.48%• | 22.65% | 14.02%• | 14.64%• | 17.21% |
| Win/Loss• | 9/3 | 10/2 | 11/1 | 11/1 | 12/0 | 11/1 | 11/1 | 9/3 | 11/1 | — |
| F-rank | 4.67 | 5.33 | 5.75 | 5.66 | 6.75 | 7.17 | 7 | 5.33 | 4.75 | **2.58** |
| P-value | 0.001 | 0.011 | 0.013 | 0.0261 | 0.0002 | 0.0007 | 0.001 | 0.021 | 0.008 | — |

\* The smaller F-rank value denotes a higher rating prediction accuracy on RD

## C. Ablation Study

We performed an ablation study to find out what is the impact of the model components on its performance. For this purpose three separate sets of experiments are done in this paper, the first one is XGBoost to predict the stochastic and seasonal components, while the trend component is predicted using ARIMA only. The second one is to predict the trend component using ARIMA-SVR and the irregular component using XGBOOST only, while the seasonal component takes the historical month. The third item is predicted for the trend component as well as the seasonal component, while the stochastic component is taken as the historical monthly average.

Fig. 6 shows that the prediction of the irregular component has a significant impact on the model, and not using XGBoost to predict it will make the model much less accurate. In addition, the prediction of the other two components also has an impact on the model but not as obvious as the prediction of the irregular component.

## V. CONCLUSIONS

This study proposes a hybrid model combined decomposition technology to accurately forecast monthly electricity load. It adopts two-fold ideas: a) The electricity load is decomposed into seasonal component, irregular component, and trend component to remove the influence of irrelevant factors on the prediction results. b) Combining the advantages of ML and statistic learning methods, while considering important relevant features to predict each component separately. The simulation study shows that the proposed model is comparable to fully trained artificial intelligence, statistical learning, and hybrid model, and is superior to them both in prediction accuracy and stability.

In future studies, we plan to introduce domain knowledge into our component prediction model, invest in the prediction of irregular components to improve prediction accuracy, and develop integrated methods.

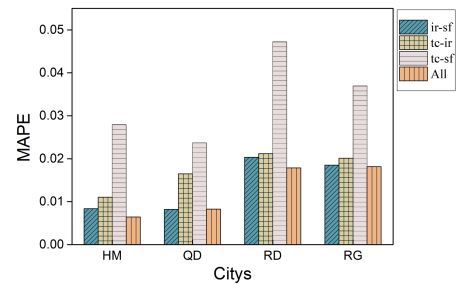

Fig. 6. The result of ablation study